\documentclass[11pt]{article}

\usepackage{dsfont}
\usepackage{relsize}
\usepackage{color}

\usepackage{textcomp}
\usepackage{url}
\usepackage{geometry}
\usepackage[cmex10]{amsmath}
\usepackage{amssymb}
\usepackage{topcapt}
\usepackage{booktabs}
\usepackage{times}
\usepackage{bm}
\usepackage{algorithmic}
\usepackage{algorithm}
\usepackage{array}
\usepackage{color}

\usepackage[outerbars]{changebar}

\usepackage[sort&compress]{natbib}
\bibpunct{[}{]}{,}{n}{,}{,}

\makeatletter
  \def\tagform@#1{\maketag@@@{(#1)\@@italiccorr}}
\makeatother

\usepackage{ifpdf}
\ifpdf
\usepackage[pdftex]{graphicx}
\usepackage{epstopdf}
\else
	\usepackage[dvips]{graphicx}
	\usepackage{color,psfrag}
\fi
\usepackage{gensymb}

\extrafloats{100}

\begin{document}

\title{\textbf{Ghost Reduction in Echo-Planar Imaging by Joint Reconstruction of Images and Line-to-Line Delays and Phase Errors}}
\author{Julianna D Ianni$^{1,2}$, E Brian Welch$^{1,3}$, William A Grissom$^{1,2,3,4}$}
\date{\vspace{-5ex}}
\maketitle
\begin{center}
$^1$Vanderbilt University Institute of Imaging Science, 
$^2$Department of Biomedical Engineering,
$^3$Department of Radiology,
$^4$Department of Electrical Engineering, Vanderbilt University, Nashville, TN, United States \\
\end{center}
\large \emph{Submitted to Magnetic Resonance in Medicine}
\section*{Abstract} 
{\bf Purpose:}
To correct line-to-line delays and phase errors in echo-planar imaging (EPI).\\
{\bf Theory and Methods:}
EPI- trajectory auto-corrected image reconstruction (EPI-TrACR) is an iterative maximum-likelihood technique that 
exploits data redundancy provided by multiple receive coils between nearby lines of k-space 
to determine and correct line-to-line trajectory 
delays and phase errors that cause ghosting artifacts.
EPI-TrACR was applied to in vivo data acquired at 7 Tesla 
across acceleration and multishot factors, and in a dynamic time series.
The method was efficiently implemented using a segmented FFT and 
compared to a conventional calibrated reconstruction.   
\\
{\bf Results:}
Compared to conventional calibrated reconstructions,
EPI-TrACR reduced ghosting up to moderate acceleration factors and across multishot factors.  
It also maintained low ghosting in a dynamic time series. 
Averaged over all cases, 
EPI-TrACR reduced root-mean-square ghosted signal outside the brain by 27\% compared to calibrated reconstruction.\\
{\bf Conclusion:} EPI-TrACR is effective in automatically correcting line-to-line delays and phase errors in multishot, 
accelerated, and dynamic EPI. 
While the method benefits from additional calibration data, it is not a requirement.\\ 
{\bf Key words:} image reconstruction; EPI; parallel imaging; phase correction; eddy currents; ghosting

\section*{Introduction}

Echo-planar imaging (EPI) is a fast imaging technique in which multiple Cartesian lines of k-space are measured per excitation. 
 It is widely used in functional magnetic resonance imaging (fMRI) and diffusion weighted imaging (DWI).
However, EPI images contain ghosting artifacts due to trajectory delays and phase errors between adjacent 
k-space lines that result from eddy currents created by rapidly switched readout gradients.

\par The most common methods to correct EPI ghosting artifacts are based on the collection of calibration data from which delays and phase shifts can be estimated and applied in image reconstruction \cite{Hu1996,Wong1992,Wan1997,Reeder1999,Chen2004,xu:mrm:2010}.
Usually this data comes from a separate acquisition  
without phase encoding gradient blips, acquired before the imaging scan.  
Corrections can also be made by re-acquiring EPI k-space data that is
offset by one k-space line so that odd k-space lines become even and vice versa \cite{Hu1996,Xiang2007}. 
The gradient impulse response function can also be measured and applied to predict errors \cite{Campbell-Washburn2014}. 
However, these methods cannot correct dynamic errors caused by effects such as gradient coil heating.
Dynamic errors can be compensated by measuring calibration data within the imaging sequence itself,
for example by reacquiring the center line of k-space within a single acquisition \cite{Jesmanowicz1993}. 
However, these approaches result in a loss of temporal resolution.
Alternatively, dynamic errors can be measured during a scan without modifying the sequence using field-probe measurements \cite{Barmet2008,Kasper2015,Wilm2016}. 
However, the hardware required to make those measurements can take up valuable space in the scanner bore and 
is not widely available at the time of writing.

\par As an alternative to separate calibration measurements, 
many retrospective methods attempt to correct ghosting based on the EPI data 
or images themselves.
The image-based methods \cite{Buonocore1997a,Buonocore2001a,Foxall1999,Lee2002}
rely on the assumption that some part of the initial image contains no ghosted signal. 
Another group of methods makes corrections based on finding phased array combinations that cancel ghosts \cite{Kellman2006,Li2013,Kim2008a,Xie2016,Hoge2010}. 
Several methods use parallel imaging to separately reconstruct images from odd and even lines and then combine them,
and these have further been combined with a dynamically alternating phase encode shift or direction \cite{Li2013,Kim2008a,hennel:mrm:2016,Xie2016,lee:mrm:2016}.
However, relying on undersampled data for calibration weights may make these approaches unstable, 
and some methods reduce temporal resolution.
Importantly, almost all these retrospective methods
are either incompatible or have not been validated with multi-shot EPI,
and most are either incompatible with parallel imaging acceleration or have only been implemented and 
validated with small acceleration factors.

\par In this work, a flexible EPI- trajectory auto-corrected image reconstruction (EPI-TrACR) 
is proposed that alleviates ghosting artifacts
by exploiting data redundancy between adjacent k-space lines in multicoil EPI data. 
It is an extension of a previously-described method for automatic non-Cartesian trajectory error correction (TrACR-SENSE) \cite{Ianni2015} 
to the joint estimation of images and line-to-line delays and phase errors in EPI. 
In the following we describe the method, 
including an efficient segmented FFT algorithm for delayed EPI k-space trajectories.
The method is then validated in vivo at 7 Tesla,
at multiple acceleration and multishot factors and in a dynamic time series.
It is demonstrated that EPI-TrACR reduces dynamic ghosting
and is compatible with multishot EPI and acceleration. 
Furthermore, the method benefits from initialization with calibration data 
but does not require it at moderate acceleration and multishot factors.

\section*{Theory}

\subsection*{Problem Formulation}

EPI-TrACR jointly estimates images, delays and phase shifts by fitting an extension of the SENSE 
MR signal model \cite{Pruessmann:1999:Magn-Reson-Med:10542355} to EPI k-space data: 
\begin{equation}
y_{c}[m,n] = \sum_{i=1}^{N_s}  e^{-\imath 2 \pi (( k^x_{m} +\Delta k^x_n ) x_i + k^y_{n} y_i)} e^{\imath \Delta \phi _n} s_{ci} f_{i} ,
\label{eq:model}
\end{equation}
where $y_{c}[m,n]$ is the signal measured in coil $c$ at the $m$th time point of the $n$th phase-encoded echo,
$k^{x}_m$ is the k-space coordinate in the readout/frequency encoded dimension and $\Delta k^x_n$ is the trajectory 
delay in that dimension for the $n$th echo (out of $N$ echoes), 
$k^{y}_n$ is the $n$th echo's k-space coordinate in the phase-encoded dimension,
$\Delta \phi_n$ is the phase shift of the $n$th echo resulting from zeroth-order eddy currents,
$s_{ci}$ is coil $c$'s measured sensitivity at $(x_i,y_i)$,
$f_i$ is the image at $(x_i,y_i)$, 
and $N_s$ is the number of pixels in the image. 
The variables in this model are the image $\bm{f}$ and the delays and phase shifts $\{(\Delta k^x_n, \Delta \phi_n)\}_{n=1}^N$,
and it is fit to measured data $\tilde{y}_{c}[m,n]$ by minimizing the sum of squared errors between the two.
This is done while constraining the delays and phase shifts  
so that a single delay and phase shift pair applies to all of a shot's odd echoes and another pair applies to all of its even echoes,
with separate parameters for each shot.
The first shot's odd echoes 
serve as a reference and are constrained to have zero
delay and phase shift. 
Overall, a total of $2 (2 N_{shot} - 1)$ delay and phase shift parameters are fit to the data along with the image.

\subsection*{Algorithm}
The EPI-TrACR algorithm minimizes the data-model error by alternately updating the estimated image $\bm{f}$, 
the k-space delays $\{\Delta k^x_n\}_{n=1}^N$, and the phase shifts $\{\Delta \phi_n\}_{n=1}^N$. 
The image is updated with a conjugate-gradient (CG) SENSE reconstruction \cite{pruess:mrm01:aise}. The delay and phase shift updates are both performed using a nonlinear Polak-Ribi{\`e}re (CG) algorithm \cite{Press:1993aa}, 
which requires computation of the derivative of the squared data-model error with respect to those parameters.This CG algorithm was chosen for its efficiency in minimizing the data-model error in similar problems; other optimization algorithms, such as gradient descent, may be applied alternatively. Denoting the sum-of-squared errors as the function $\Psi$, 
the derivative with respect to each delay $\Delta k^x_n$ is:
\begin{equation}
\frac{\partial \Psi}{\partial \Delta k^x_n} = \sum_{c=1}^{N_c} \sum_{m=1}^{M} \sum_{i = 1}^{N_s} \Re \left \{ -\imath 2\pi x_i  e^{-\imath \Delta \phi _n} e^{\imath 2 \pi (( k^x_{m} +\Delta k^x_n ) x_i + k^y_{n} y_i)} s_{ci}^* f_i^* r_{cmn} \right \},
\label{eq:delayderiv}
\end{equation}
and the derivative with respect to each phase shift $\Delta \phi_n$ is:
\begin{equation}
\frac{\partial \Psi}{\partial \Delta \phi_n} = \sum_{c=1}^{N_c} \sum_{m=1}^{M} \sum_{i = 1}^{N_s} \Re \left \{ \imath e^{-\imath \Delta \phi _n} e^{\imath 2 \pi (( k^x_{m} +\Delta k^x_n ) x_i + k^y_{n} y_i)} s_{ci}^* f_i^* r_{cmn} \right \},
\label{eq:phsderiv}
\end{equation} 
where $\Re$ denotes the real part, $^*$ is complex conjugation, and $r_{cmn}$ is the residual error between the measured data and the model given the current parameter estimates, $\hat{\bm{f}}$, $\Delta \hat{k}^x_n$, and $\Delta \hat{\phi}_n$:
\begin{equation}
r_{cmn} = \tilde{y}_{c}[m,n] - \sum_{i=1}^{N_s}  e^{-\imath 2 \pi (( k^x_{m} +\Delta \hat{k}^x_n ) x_i + k^y_{n} y_i)} e^{\imath \Delta \hat{\phi}_n} s_{ci} \hat{f}_{i}.
\label{eq:resid}
\end{equation}
To constrain the delays and phase shifts to be the same for the set of odd or even echoes of each shot, 
the derivatives above are summed across the echoes in that set,
and a single delay and shift pair is determined for the set each CG iteration. 
The updates are alternated until the data-model error stops changing significantly.

\subsubsection*{Segmented FFTs}
\begin{figure}[ht]
\centering
\includegraphics[width=\textwidth]{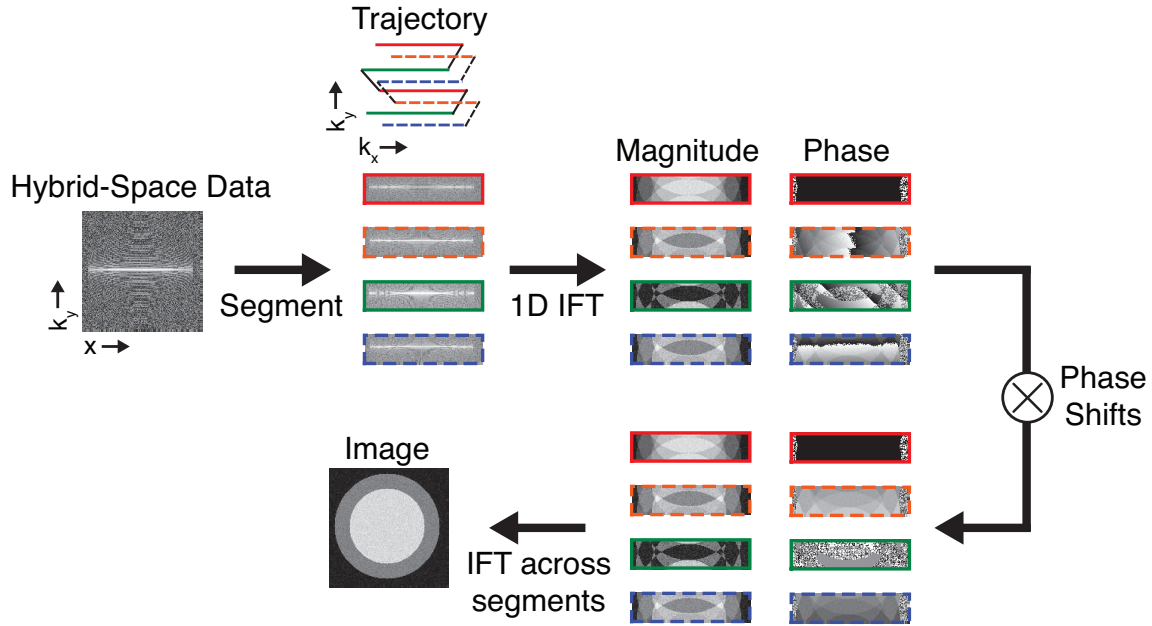}
\caption{Illustration of the inverse segmented FFT, 
starting with 2-shot $x$-$k_y$ EPI data corrupted by line-to-line delays and phase errors.
First the data are segmented into 2$N_{shot}$ submatrices and individually inverse Fourier transformed.
Then each image-domain submatrix is phase shifted to account for its offset in $k_y$, its phase error, and its delay. 
Finally, an inverse Fourier transform is calculated across the segments, and the result is reshaped into the image.}
\label{fig:segfft}
\label{basisfig}
\end{figure}

Since a delayed EPI trajectory is non-Cartesian, the model in Equation \ref{eq:model} corresponds to a 
non-uniform discrete Fourier transform (DFT) of the image.
Non-uniform fast Fourier transform (FFT) algorithms (e.g., Ref. \cite{Fessler:2003dz}) are typically used to efficiently evaluate non-uniform DFTs,
but they use gridding, which would result in long compute times in EPI-TrACR,
since Equation \ref{eq:model} is repeatedly evaluated by the algorithm. 
Figure \ref{fig:segfft}
 illustrates a segmented FFT algorithm that applies the delays as phase ramps in the image 
domain, instead of gridding the delayed data in the frequency domain. 
In addition to eliminating gridding, 
this also enables the data to be FFT'd in the frequency-encoded dimension before starting EPI-TrACR, 
so that the algorithm only needs to compute 1D FFTs in the phase-encoded dimension. 
The figure shows an inverse segmented FFT (k-space to image space) for a 2-shot dataset with delays and phase shifts, 
which comprises the following steps:
\begin{enumerate}
\item The data in each set of odd or even echoes of each shot are collected into $2N_{shot}$ 
submatrices of size $M \times (N/(2 \times N_{shot}))$, 
and the 1D inverse FFT of each submatrix is computed in the phase-encoded dimension.  
\item The estimated phase shifts are applied to each submatrix.
\item A phase ramp is applied in the phase-encoded spatial dimension of each submatrix 
to account for that set's relative position in the phase-encoded k-space dimension. 
This is necessary since the inverse FFTs assume all the submatrices are centered in k-space.
\item The phase ramp corresponding to each set's estimated delay is applied to its submatrix in the frequency-encoded spatial dimension. 
\item For each submatrix entry, the inverse DFT across submatrices is computed to obtain $2N_{shot}$ subimages of size $M \times (N/(2 \times N_{shot}))$, which are concatenated in the column dimension to form the final $M \times N$ image. 
\end{enumerate}
For efficiency, 
the phase shifts of steps 2 through 4 are combined into a single precomputed matrix that is applied to each 
submatrix by elementwise multiplication.
To perform the forward segmented FFT (image space to k-space), the steps are reversed, with the phase ramps and shifts negated. 
Steps 1 and 5 dominate the computational cost, and respectively require $O\left(M N N_{shot}\right)$ and $O\left(M N \log\left(N/\left(2N_{shot}\right)\right) \right)$ operations. 

\section*{Methods}
\subsection*{Algorithm Implementation}
The EPI-TrACR algorithm was implemented in MATLAB 2016a (The Mathworks, Natick, MA, USA) on a workstation with 
dual 6-core 2.8 GHz X5660 Intel Xeon CPUs (Intel Corporation, Santa Clara, CA) and 96 GB of RAM. 
For each iteration of the algorithm's outer loop, 
image updates were initialized with zeros to prevent noise amplification,
and were performed using MATLAB's \texttt{lsqr} function and a fixed tolerance of $10^{-1}$, 
capped at 25 iterations.
CG delay and phase updates were each fixed to a maximum of 5 iterations per outer loop iteration, 
and terminated early if all steps were less than $10^{-6}$ cm$^{-1}$ (for delays) or $10^{-6}$ radians (for phase shifts).
The maximum permitted delay in a single iteration was limited to $1/FOV$, and the maximum permitted phase step in a single iteration was limited to $\pi/10$ radians.
Outer loop iterations stopped when the change in squared error was less than the previous iteration's error times $10^{-6}$. 
Code and example data for EPI-TrACR can be downloaded at \url{https://bitbucket.org/wgrissom/tracr}.

\subsection*{Experiments}
A healthy volunteer was scanned on a 7T Philips Achieva scanner (Philips Healthcare, Best, Netherlands) 
with the approval of the Institutional Review Board at Vanderbilt University. 
A birdcage coil was used for excitation and a 32-channel head coil for reception (Nova Medical Inc., Wilmington, MA, USA). 
EPI scans were acquired with 24 $\times$ 24 cm FOV, 1.5 $\times$ 1.5 $\times$ 3 mm$^3$ voxels, 
TR 3000 ms, TE 56 ms, flip angle 60$\degree$. 
They were repeated for 1 to 4 shots, 
acceleration factors of 1x to 4x, 
and a single scan (2-shot, 1x) was performed with 20 repetitions. 
The TE of 56 ms was chosen to maintain the same contrast between images, and was the shortest possible for the single-shot, 1x acquisition, which had a readout duration of 102 ms.
A calibration scan with phase encodes turned off was acquired in each configuration,
and delays and phase shifts were estimated from it using cross-correlation followed by an optimization transfer-based 
refinement \cite{Funai2008}.
SENSE maps were also collected using the vendor's mapping scan. 
Images were reconstructed to 160 $\times$ 160 matrices using \texttt{lsqr} with no corrections, 
conventional calibration (using the phase and delay estimates from the calibration scan),
EPI-TrACR initialized with the delays and phase shifts from the conventional calibration,
and EPI-TrACR initialized with zeros. For comparison of EPI-TrACR corrections on the time series data with another dynamic method, PAGE \cite{Kellman2006} was also implemented.
To characterize the amount of data necessary for the EPI-TrACR reconstruction, 
the algorithm was repeated after truncating the 2-shot, 
1x in vivo data in both k-space dimensions across a range of truncation factors. 
The reconstructed image resolution within EPI-TrACR was correspondingly reduced in each case, 
so that the image matrix size matched the data matrix size. 
The final estimated delays and phase shifts were then applied in a full-resolution reconstruction.
Except where indicated, 
displayed images shown are windowed down to 20\% of their maximum amplitude for clear display of ghosting,
and ghosted signals were measured in all images as the root-mean-square (RMS) signal 
outside an elliptical region-of-interest that excluded the brain and skull. 


\section*{Results}
\begin{figure}
\centering
\includegraphics[width=\textwidth]{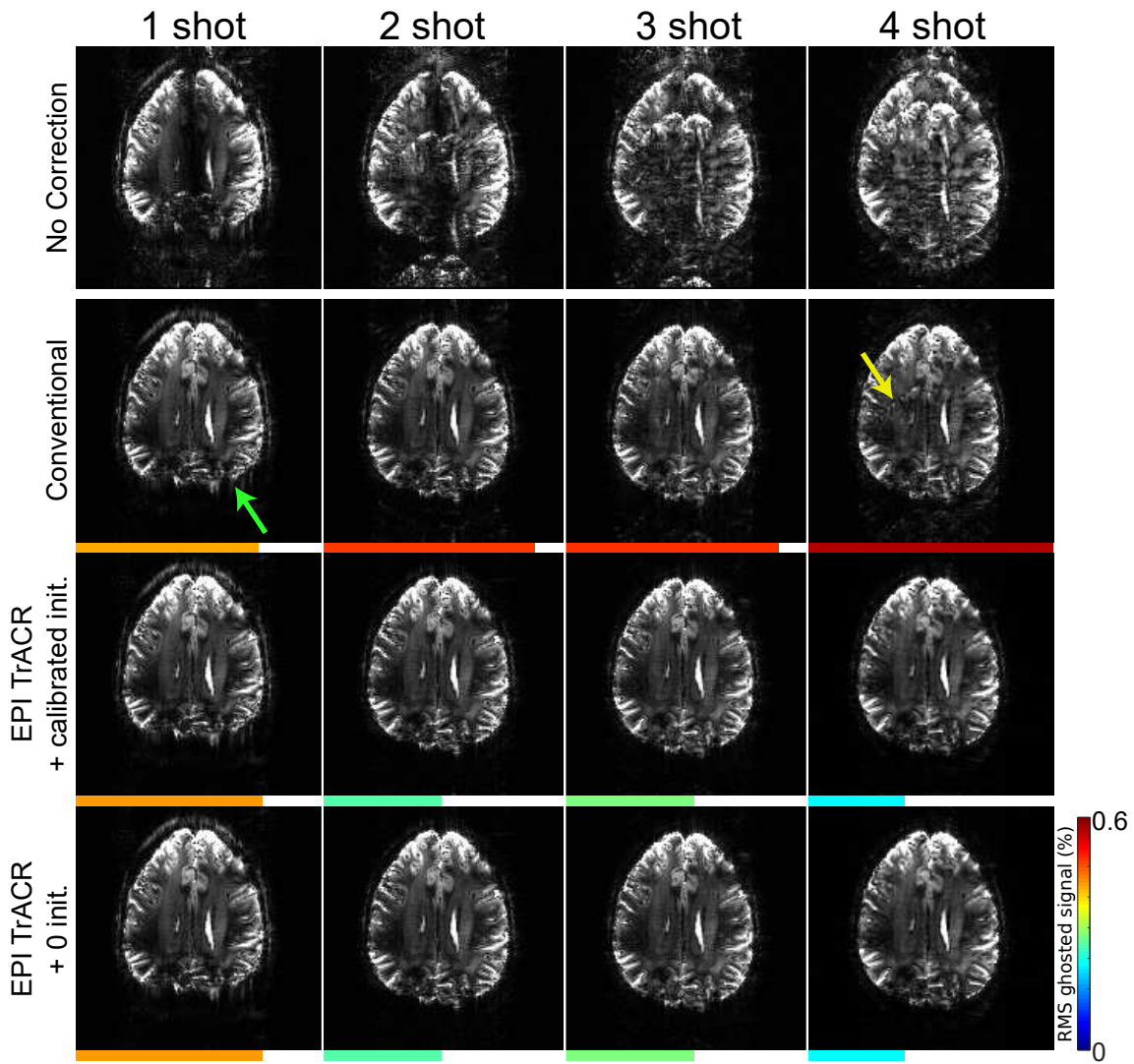}
\caption{Multishot echo-planar images (no acceleration) reconstructed with no correction, 
conventional calibrated correction, 
EPI-TrACR with calibrated initialization, 
and EPI-TrACR with zero initialization. 
The length and color of the horizontal bars beneath each image represent the residual RMS ghosted signal 
as a percentage of maximum image intensity, as defined by the color scale on the right. The green arrow in the conventional single-shot reconstruction indicates an off-resonance artifact which appears in all of the single-shot reconstructions.
The yellow arrow in the conventional 4-shot reconstruction indicates an edge that aliased into the brain, which is not visible in the EPI-TrACR reconstructions.}
\label{fig:multishot}
\end{figure}

\begin{figure}[!ht]
\centering
\includegraphics[width=\textwidth]{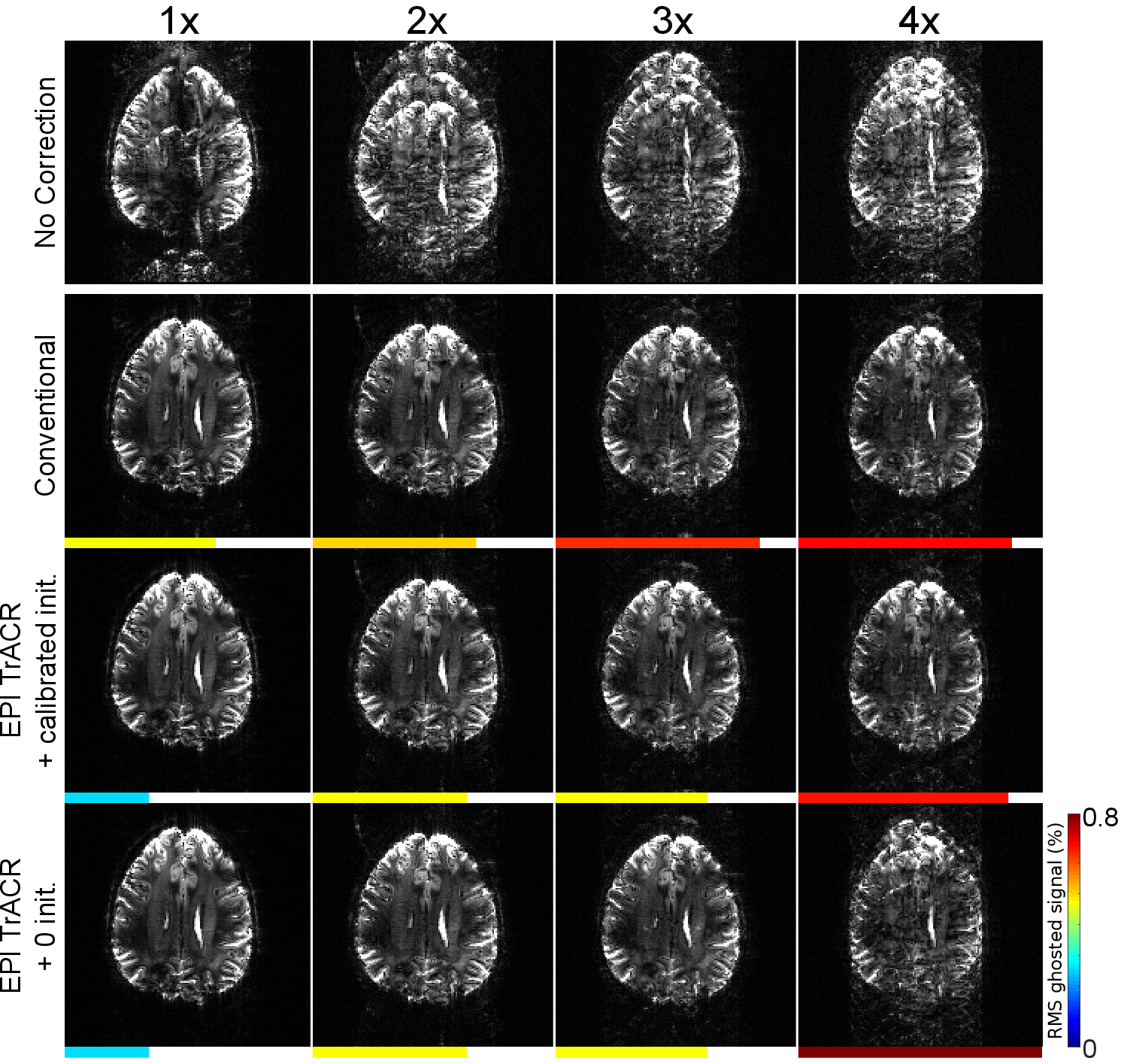}
\caption{1x, 2x, 3x and 4x 2-shot echo-planar images reconstructed with no correction, 
conventional calibrated correction, 
EPI-TrACR with calibrated initialization, 
and EPI-TrACR with zero initialization. 
The length and color of the horizontal bars beneath each image represent the residual RMS ghosted signal 
as a percentage of maximum image intensity, as defined by the color scale on the right.}
\label{fig:acc}
\end{figure}

\begin{figure}[!ht]
\centering
\includegraphics[width=\textwidth]{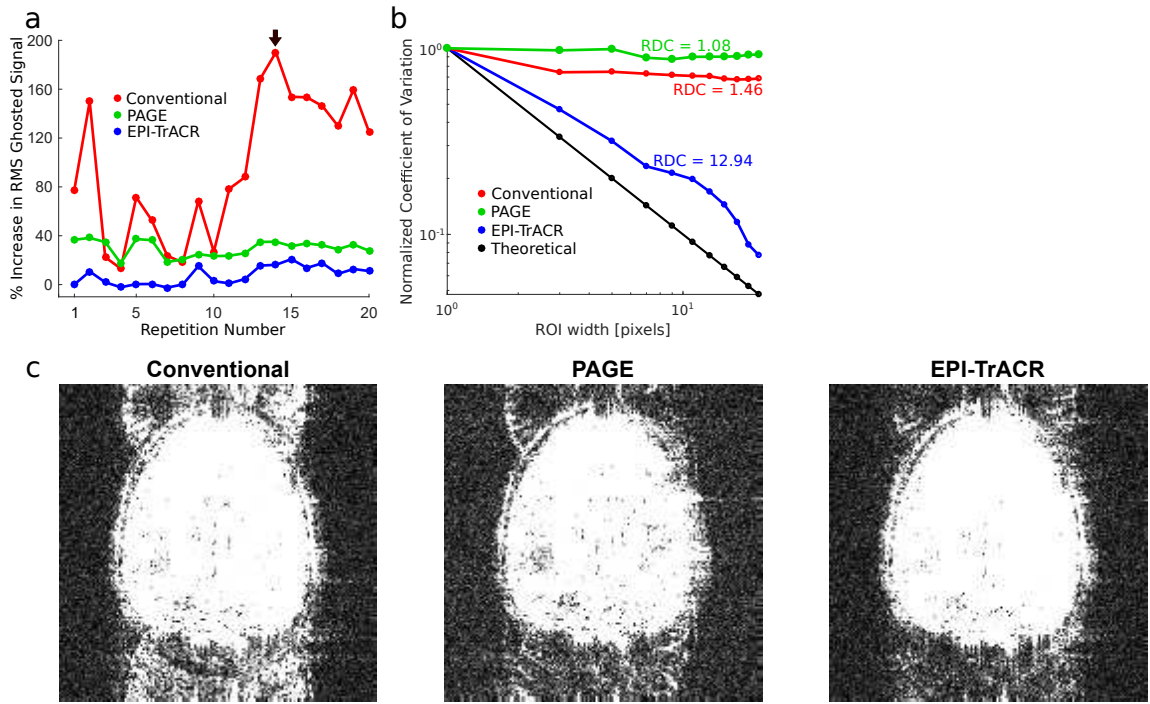}
\caption{2-shot echo-planar images over 20 repetitions reconstructed using conventional calibration, 
PAGE, and EPI-TrACR with zero initialization. 
(a) Percentage increase in RMS ghosted signal versus repetition number, 
normalized to that of the EPI-TrACR reconstruction of the first repetition. 
(b) Weisskoff plot showing the normalized coefficient of variation over repetitions for an ROI of increasing size, 
for conventional calibration, PAGE, and EPI-TrACR compared to the theoretical ideal.
(c) Windowed-down conventional calibration, PAGE, and EPI-TrACR reconstructions,
at the 14th repetition (indicated by the arrow in (a)).}
\label{fig:dyns}
\end{figure}

Figure \ref{fig:multishot}
 shows reconstructed images across multishot factors.
Ghosting was lowest with EPI-TrACR in all cases, 
and the differences between zero initialization and calibrated initialization results are negligible:
averaged across multishot factors, 
the RMS difference between estimated delays and phase shifts with and without calibrated initialization was 0.014\%.
Compared to conventional calibration-based correction, 
EPI-TrACR RMS ghosted signals were on average 37\% lower.
In addition, the conventional 4-shot reconstruction contained a visible aliased edge inside the brain (indicated by the yellow arrow),
which did not appear in the EPI-TrACR reconstructions. 
All of the single-shot reconstructions contain a visible off-resonance artifact at the back of the brain (indicated in the conventional reconstruction by the green arrow).
Figure \ref{fig:acc} 
shows reconstructed 2-shot EPI images with 1-4$\times$ acceleration.
Compared to conventional calibration, 
EPI-TrACR with calibrated initialization again reduced ghosting up to 4$\times$ acceleration, 
and RMS ghosted signals were 18\% lower on average. 
Furthermore, 
EPI-TrACR estimates matched with and without calibrated initialization up to 3$\times$ acceleration:
averaged across factors of 1-3$\times$, 
the RMS difference between estimated delays and phase shifts with and without calibrated initialization was 0.024\%.
Figure \ref{fig:dyns}a
 plots RMS ghosted signal across repetitions of the 2-shot/1$\times$ scan for conventional calibrated reconstruction,
PAGE, and EPI-TrACR.
The signal levels are normalized to that of the first repetition's EPI-TrACR reconstruction.
Figure \ref{fig:dyns}b shows a Weisskoff plot \cite{Weisskoff1996} for all three reconstructions compared to the theoretical ideal; 
the coefficient of variation over repetitions is plotted for an ROI of increasing size.
Figure \ref{fig:dyns}c shows conventional calibration, PAGE, and EPI-TrACR (with zero initialization)
images at the 14th repetition.  
The conventional, PAGE, and EPI-TrACR images at the 14th repetition respectively have 190\%, 35\%, and 16\% higher RMS ghosted signal compared to 
the first repetition EPI-TrACR reconstruction.
EPI-TrACR maintained consistently low ghosting across repetitions, 
and a much higher radius of decorrelation than the conventional
calibrated and PAGE reconstructions.  
A video of the full time series is provided as Supporting Information.

\par
\begin{figure}[!ht]
\centering
\includegraphics[width=5in]{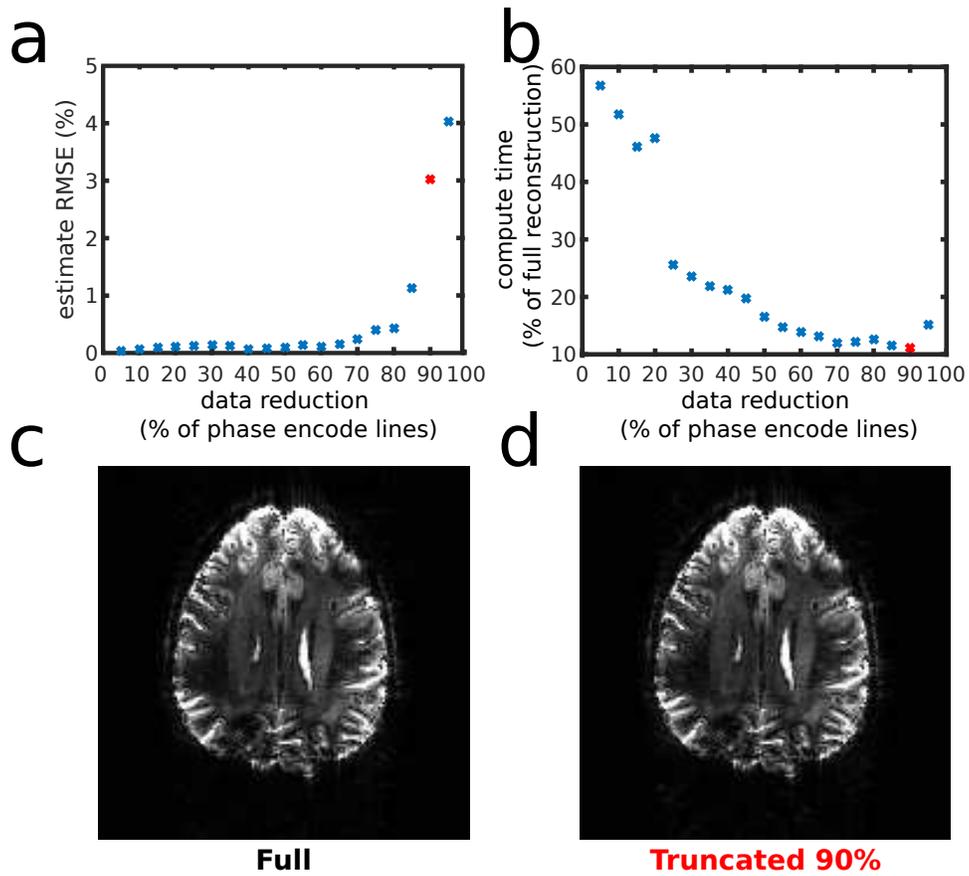}
\caption{2-shot, 1x-accelerated EPI-TrACR reconstructions from truncated data. 
The plots show (a) combined root mean square error (RMSE) in the estimates of DC and linear phase shifts compared to full-data EPI-TrACR estimates and (b) compute time as a percentage of a full-data EPI-TrACR compute time; 
both are shown as a function of percentage of degree of data reduction in each dimension. 
(c-d) Images reconstructed using the full data EPI-TrACR estimates using 90\%-truncated EPI-TrACR estimates (16 PE lines) (d) (the red data point in a-b).}
\label{fig:trunc}
\end{figure}

 The truncated 2-shot EPI-TrACR results are shown in Figure \ref{fig:trunc}. 
Figure \ref{fig:trunc}a shows that delay and phase shift estimation errors relative to full-data EPI-TrACR estimates
are low up to very high truncation factors,
and Figure \ref{fig:trunc}b shows that compute time can be reduced up to 90\% by truncating the data by 90\%.
Figures \ref{fig:trunc}c and d show that images reconstructed with full data and 
90\%-truncated data delay and phase estimates are indistinguishable: 
RMS ghosted signal was 8\% higher in the truncated EPI-TrACR image versus the full-data reconstruction,
but still 40\% lower than the conventional calibrated reconstruction (which appears in Figure \ref{fig:multishot}). 
For greater than 90\% truncation though, the compute time starts to increase again due to increasing iterations.
For full data, reconstruction times ranged from one minute (for 1 shot, 1$\times$ acceleration, and calibrated initialization)
to 88 minutes (for 2 shots, 4$\times$ acceleration, and zero initialization). 
Reconstructions using NUFFTs \cite{Fessler:2003dz} in place of the piecewise FFTs in EPI-TrACR ranged from 8 minutes 
(for 1 shot, 1$\times$ acceleration, and calibrated initialization) to 269 minutes 
(for 2 shots, 4$\times$ acceleration, and zero initialization).


\section*{Discussion}

EPI-TrACR is an iterative algorithm that jointly estimates EPI echo delays and phase shifts, 
along with images that are compensated for them. 
Compared to conventional calibrated corrections, 
EPI-TrACR consistently reduced image ghosting across multishot factors, acceleration factors, and a dynamic time series.
In most cases it was able to do so without being initialized with calibrated delays and phase shifts.
A further characterization of the convergence of EPI-TrACR with varying initialization is included as Figure \ref{fig:initialization_error} in the Supporting Information.
An additional validation experiment comparing EPI-TrACR 
estimates (1 shot, 1$\times$ acceleration) to a full k-space trajectory measurement \cite{Welch2017} in a phantom at 3 Tesla is shown in Supporting information Figure S2. 
The EPI-TrACR bulk line delay estimate was similar to the median measured delay (13\% RMS difference), 
and the EPI-TrACR image contained 19\% lower RMS ghosted signal.
Because EPI-TRACR relies on data redundancy between nearby lines of k-space,
its performance is expected to degrade with increasing acceleration factor, which was observed here. 
Nevertheless, when initialized with calibrated delays and phase shifts, 
the method always reduced ghosting compared to conventional calibrated reconstruction. 
\par The main tradeoff for EPI-TrACR's improved delay and phase shift estimates is increased computation,
but this can be mitigated in several ways. 
First, we showed that compute time can be reduced by truncating the data matrix down to the low frequencies, 
without compromising the delay and phase shift estimates. 
Compute times are also shorter when the algorithm is initialized with calibrated estimates, 
since fewer iterations are required to reach a solution. 
The algorithm could be applied in parallel across repetitions or slices, 
or within the algorithm the FFTs could be parallelized across receive coils.

\par There are a number of ways the method could be extended.
First, in the present work it was assumed that all the echoes within a set of even or odd echoes of a shot had the same 
delay and phase shift.
However, it is also possible to estimate different delays and phase shifts for different echoes within a set 
by expressing them as a weighted sum of basis functions. 
We have previously tested this extension using triangular basis functions,
but found little improvement with our data. 
Nevertheless, as others may find it useful this functionality is included in the provided code.
Second, the method could be extended to jointly estimate a single set of delays and phase shifts over a whole
stack of slices simultaneously, which would increase the effective signal-to-noise ratio for estimation. 
This could in particular help for highly accelerated acquisitions where the method is currently more sensitive to poor initialization.

\section*{Conclusions}
The EPI-TrACR
method alleviates ghosting artifacts
by exploiting data redundancy between adjacent k-space lines in multicoil EPI data. 
It benefits from initialization with calibration data but does not require it at moderate acceleration and multishot factors.
EPI-TrACR reduced dynamic ghosting without sacrificing temporal resolution,
is compatible with multishot and accelerated acquisitions, 
and unlike previous data-based approaches, it does not rely on a ghost-free image region. 
It was validated in vivo at 7T, 
at multiple acceleration and multishot factors and in a dynamic time series.

\section*{Acknowledgments}

This work was supported by NIH grants R25 CA136440, R01 EB016695, and R01 DA019912.
The authors would like to thank Dr. Manus Donahue for help with experiments. 
\pagebreak

\section*{Supporting Information Figures and Captions}

\renewcommand\thefigure{S\arabic{figure}} 
\setcounter{figure}{0}  
\begin{figure}[!ht]
\centering
\includegraphics[width=6in]{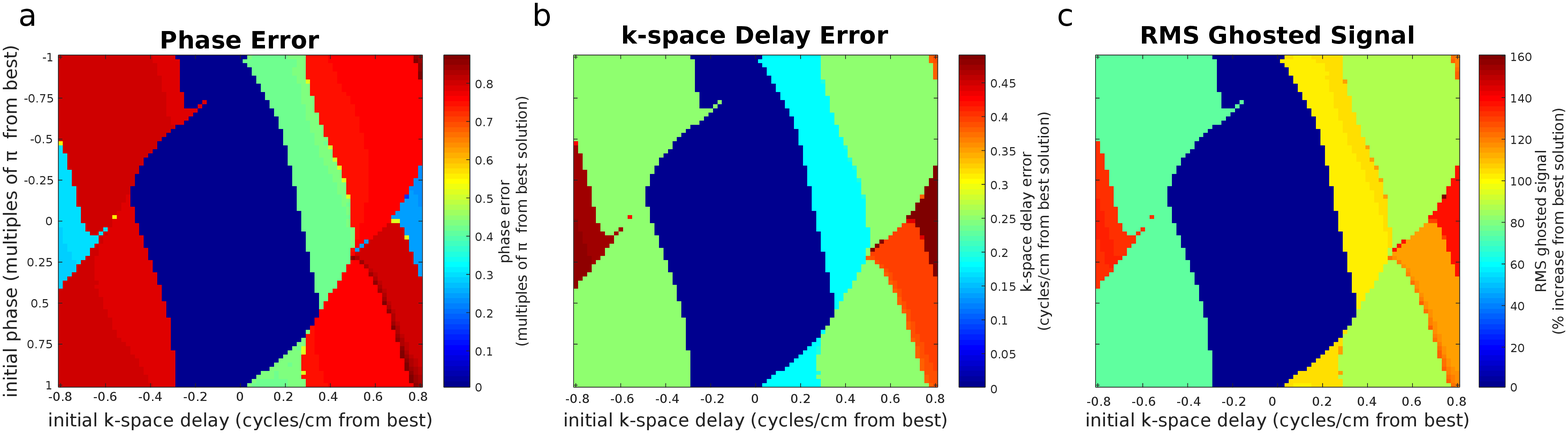}
\caption{The performance of EPI-TrACR on the single-shot, 1x data is characterized for varying combinations of initialization parameters (i.e. a single phase offset and a single k-space delay for each). Shown are the resulting final
(a) phase error (multiples of $\pi$),
(b) k-space delay error (cycles/cm), and
(c) RMS ghosted signal, for each initialization.
All parameters and errors are expressed relative to the solution with the lowest ghosting ("best" solution), which corresponds to a phase offset of -2.96 radians and a k-space delay of  0.075 cycles/cm. 
}
\label{fig:initialization_error}
\end{figure}
\begin{figure}[!ht]
\begin{center}
\includegraphics[width=5in]{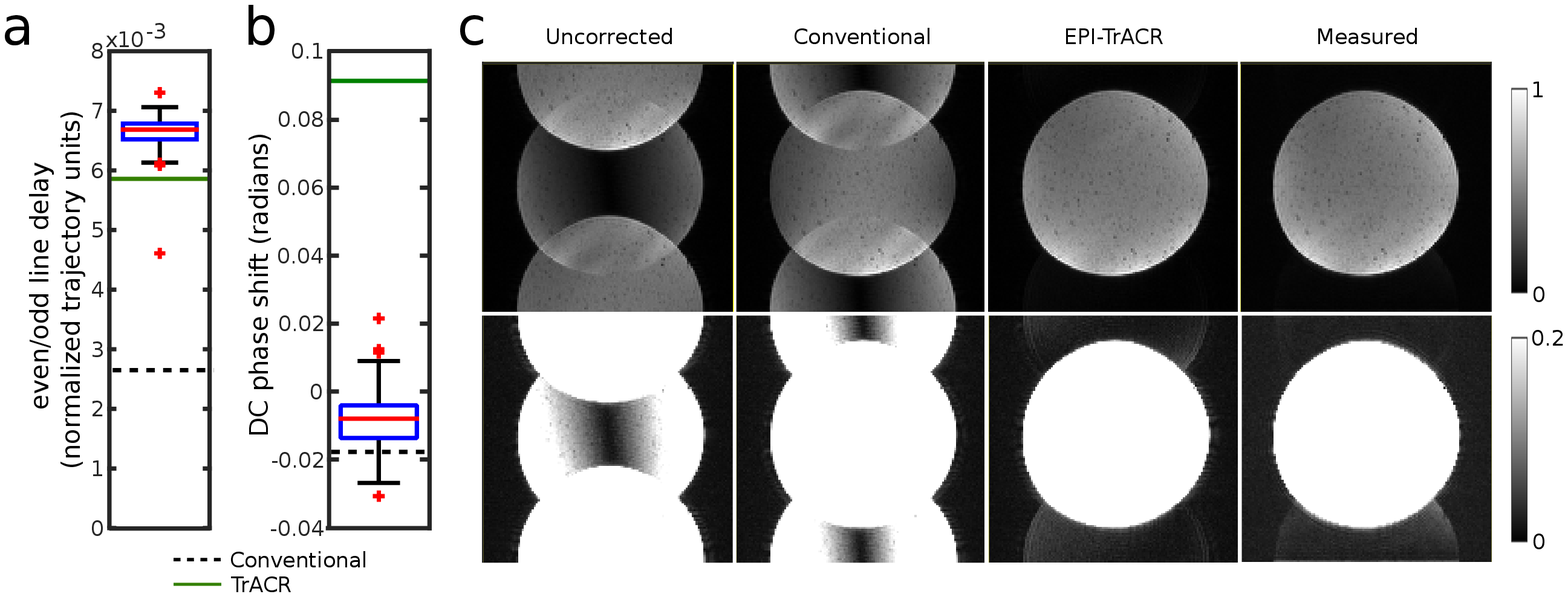}
\end{center}
\end{figure}

\noindent Figure S2: A separate experiment was performed in a phantom at 3T (Philips Achieva), using a volume coil for excitation and a 32-channel coil for reception (Nova Medical Inc., Wilmington, MA, USA). 
Data were collected for a single off-axis slice (5$\degree$/20$\degree$/30$\degree$) using a single-shot EPI scan with 60 dynamics; scan parameters were: 23 $\times$ 23 cm FOV, 1.8 $\times$ 1.8 $\times$ 4 mm voxels, TR 2000 ms, TE 43 ms, flip angle 90$\degree$. 
The trajectory was measured for a single dynamic using a modified Duyn method \cite{Gurney2005,Welch2017}. 
A SENSE map and a calibration scan with phase encodes turned off were also collected as for the in vivo data. 
From the measured trajectory, 
trajectory delays were estimated as the average shift between each pair of odd and even lines over the middle quarter of the readout dimension. 
EPI-TrACR was used to reconstruct the phantom data in the same manner as for the in vivo data described above. 
Residual ghosted signal was calculated for all images as the root-mean-square (RMS) signal outside an elliptical region-of-interest masking out the phantom. 
Shown in this figure are boxplots of the measured line-to-line trajectory delays in the readout dimension (a) and DC phase errors (b), with lines superimposed to mark the conventional (dashed black) and EPI-TrACR (solid green) estimates. (c) Corresponding images shown are conjugate-gradient (CG) reconstructions of the first dynamic of phantom data along the uncorrected trajectory, that corrected by conventional estimates, that estimated with EPI-TrACR (with calibrated initialization), and the measured EPI trajectory. Images are shown at full magnitude (top) and windowed to 20\% (bottom). RMS image ghosting is 19\% lower in the TrACR image than in the measured image. The bulk even/odd line shift estimated was approximately 13\% different between the two trajectories. The conventional reconstruction method was unable to adequately correct for the large amount of ghosting in the uncorrected image. Both EPI-TrACR and measured-trajectory reconstructions reduced ghosting over the conventional method. This provides additional confidence in the EPI-TrACR estimates beyond that provided by the conventional reconstruction. Residual ghosting apparent in both measured-trajectory and EPI-TrACR reconstructions may be attributed in part to the off-axis slice, which yielded particularly large trajectory and line-to-line phase shifts. The measured trajectory accounted for additional errors (e.g., a small readout-dimension shrinkage of the trajectory extent) that are not encompassed by the EPI-TrACR basis functions implemented herein; however, these measured errors do not appear to greatly improve the image over the EPI-TrACR reconstruction.

\renewcommand{\figurename}{Supporting Video}
\renewcommand\thefigure{V\arabic{figure}} 
\setcounter{figure}{0}  
\begin{figure}
\caption{2-shot echo-planar images over 20 repetitions reconstructed using conventional calibration, PAGE, and EPI-TrACR.   Shown are windowed-down conventional calibration, PAGE, and EPI-TrACR reconstructions at each dynamic (right), and corresponding percentage increase in RMS ghosted signal plotted versus repetition number, normalized to that of the EPI-TrACR reconstruction of the first repetition (left).}
\label{fig:dynamicvideo}
\end{figure}

\bibliographystyle{cse}
\bibliography{EPI_TrACR_paper}

\begin{thebibliography}{10}
\providecommand{\url}[1]{\texttt{#1}}
\providecommand{\urlprefix}{URL }

\bibitem{Hu1996}
Hu X, Le TH.
\newblock {Artifact reduction in EPI with phase-encoded reference scan}.
\newblock Magn Reson Med 1996;\hspace{0pt}36:166--171.

\bibitem{Wong1992}
Wong EC.
\newblock {Shim Insensitive Phase Correction for EPI Using a Two Echo Reference
  Scan}.
\newblock In Proc Int Soc Magn Res Med. volume~S2, 1992;\hspace{0pt} p. 4514.

\bibitem{Wan1997}
Wan X, Gullberg GT, Parker DL, Zeng GL.
\newblock {Reduction of geometric and intensity distortions in echo-planar
  imaging using a multireference scan}.
\newblock Magn Reson Med 1997;\hspace{0pt}37:932--942.

\bibitem{Reeder1999}
Reeder SB, Faranesh AZ, Atalar E, McVeigh ER.
\newblock {A novel object-independent 'balanced' reference scan for echo-planar
  imaging}.
\newblock J Magn Reson Imaging 1999;\hspace{0pt}9:847--852.

\bibitem{Chen2004}
Chen NK, Wyrwicz AM.
\newblock {Removal of EPI Nyquist ghost artifacts with two-dimensional phase
  correction}.
\newblock Magn Reson Med 2004;\hspace{0pt}51:1247--1253.

\bibitem{xu:mrm:2010}
Xu D, King KF, Zur Y, Hinks RS.
\newblock Robust {2D} phase correction for echo planar imaging under a tight
  field-of-view.
\newblock Magn Reson Med 2010;\hspace{0pt}64:1800--1813.

\bibitem{Xiang2007}
Xiang QS, Ye FQ.
\newblock {Correction for geometric distortion and N/2 ghosting in EPI by phase
  labeling for additional coordinate encoding (PLACE)}.
\newblock Magn Reson Med 2007;\hspace{0pt}57:731--741.

\bibitem{Campbell-Washburn2014}
Campbell-Washburn A, Xue H, Lederman R, Faranesh A, Hansen M.
\newblock {Real-time distortion correction of spiral and echo planar images
  using the gradient system impulse response function}.
\newblock Magn Reson Med 2016;\hspace{0pt}75:2278--2285.

\bibitem{Jesmanowicz1993}
Jesmanowicz A, Wong EC, Hyde JS.
\newblock {Phase correction for EPI using internal reference lines}.
\newblock In Proc Int Soc Magn Reson Med. 1993;\hspace{0pt} p. 1239.

\bibitem{Barmet2008}
Barmet C, {De Zanche} N, Pruessmann KP.
\newblock {Spatiotemporal magnetic field monitoring for MR.}
\newblock Magn Reson Med 2008;\hspace{0pt}60:187--197.

\bibitem{Kasper2015}
Kasper L, Bollmann S, Vannesjo SJ, Gross S, Haeberlin M, Dietrich BE,
  Pruessmann KP.
\newblock {Monitoring, analysis, and correction of magnetic field fluctuations
  in echo planar imaging time series}.
\newblock Magn Reson Med 2014;\hspace{0pt}409:396--409.

\bibitem{Wilm2016}
Wilm BJ, Dietrich BE, Reber J, Vannesjo SJ, Pruessmann KP.
\newblock {Gradient response harvesting for continuous system characterization
  during MR sequences}.
\newblock Proc Int Soc Magn Reson Med 2016;\hspace{0pt}pp. 8--11.

\bibitem{Buonocore1997a}
Buonocore MH, Gao L.
\newblock {Ghost artifact reduction for echo planar imaging using image phase
  correction}.
\newblock Magn Reson Med 1997;\hspace{0pt}38:89--100.

\bibitem{Buonocore2001a}
Buonocore MH, Zhu DC.
\newblock {Image-based ghost correction for interleaved EPI}.
\newblock Magn Reson Med 2001;\hspace{0pt}45:96--108.

\bibitem{Foxall1999}
Foxall DL, Harvey PR, Huang J.
\newblock {Rapid iterative reconstruction for echo planar imaging.}
\newblock Magn Reson Med 1999;\hspace{0pt}42:541--7.

\bibitem{Lee2002}
Lee KJ, Barber DC, Paley MN, Wilkinson ID, Papadakis NG, Griffiths PD.
\newblock {Image-based EPI ghost correction using an algorithm based on
  projection onto convex sets (POCS)}.
\newblock Magn Reson Med 2002;\hspace{0pt}47:812--817.

\bibitem{Kellman2006}
Kellman P, McVeigh ER.
\newblock {Phased array ghost elimination}.
\newblock NMR Biomed 2006;\hspace{0pt}19:352--361.

\bibitem{Li2013}
Li H, Fox-Neff K, Vaughan B, French D, Szaflarski JP, Li Y.
\newblock {Parallel EPI artifact correction (PEAC) for N/2 ghost suppression in
  neuroimaging applications}.
\newblock J Magn Reson Imaging 2013;\hspace{0pt}31:1022--1028.

\bibitem{Kim2008a}
Kim YC, Nielsen JF, Nayak KS.
\newblock {Automatic correction of Echo-Planar Imaging (EPI) ghosting artifacts
  in real-time interactive cardiac MRI using sensitivity encoding}.
\newblock J Magn Reson Imaging 2008;\hspace{0pt}27:239--245.

\bibitem{Xie2016}
Xie VB, Lyu M, Wu EX.
\newblock {EPI Nyquist ghost and geometric distortion correction by two-frame
  phase labeling}.
\newblock Magn Reson Med 2016;\hspace{0pt}00:1--13.

\bibitem{Hoge2010}
Hoge WS, Tan H, Kraft RA.
\newblock {Robust EPI Nyquist ghost elimination via spatial and temporal
  encoding}.
\newblock Magn Reson Med 2010;\hspace{0pt}64:1781--1791.

\bibitem{hennel:mrm:2016}
Hennel F, Buehrer M, von Deuster C, Seuven A, Pruessmann KP.
\newblock {SENSE reconstruction for multiband EPI including slice-dependent N/2
  ghost correction}.
\newblock Magn Reson Med 2015;\hspace{0pt}76:873--879.

\bibitem{lee:mrm:2016}
Lee J, Jin KH, Ye JC.
\newblock Reference-free single-pass {EPI Nyquist} ghost correction using
  annihilating filter-based low rank {Hankel} matrix ({ALOHA}).
\newblock Magn Reson Med 2016;\hspace{0pt}76:1775--1789.

\bibitem{Ianni2015}
Ianni JD, Grissom WA.
\newblock Trajectory auto-corrected image reconstruction.
\newblock Magn Reson Med 2016;\hspace{0pt}76:757--768.

\bibitem{Pruessmann:1999:Magn-Reson-Med:10542355}
Pruessmann KP, Weiger M, Scheidegger MB, Boesiger P.
\newblock {SENSE: Sensitivity encoding for fast MRI}.
\newblock Magn Reson Med 1999;\hspace{0pt}42:952--962.

\bibitem{pruess:mrm01:aise}
Pruessmann KP, Weiger M, B\"{o}rnert P, Boesiger P.
\newblock {Advances in sensitivity encoding with arbitrary k-space
  trajectories}.
\newblock Magn Reson Med 2001;\hspace{0pt}46:638--651.

\bibitem{Press:1993aa}
Press WH.
\newblock {Numerical recipes in C}.
\newblock Cambridge University Press, 2nd ed., v edition, 1993.

\bibitem{Fessler:2003dz}
Fessler JA, Sutton BP.
\newblock {Nonuniform fast fourier transforms using min-max interpolation}.
\newblock IEEE Trans Signal Process 2003;\hspace{0pt}51:560--574.

\bibitem{Funai2008}
Funai AK, Fessler JA, Yeo DTB, Olafsson VT, Noll DC.
\newblock {Regularized field map estimation in MRI}.
\newblock IEEE Trans Med Imaging 2008;\hspace{0pt}27:1484--1494.

\bibitem{Weisskoff1996}
Weisskoff RM.
\newblock {Simple measurement of scanner stability for functional NMR imaging
  of activation in the brain.}
\newblock Magn Reson Med 1996;\hspace{0pt}36:643--645.

\bibitem{Welch2017}
Welch EB, Harkins KD.
\newblock {Robust k-space trajectory mapping with data readout concatentation
  and automated phase unwrapping reference point identification}.
\newblock Proc Int Soc Magn Reson Med 2017;\hspace{0pt}p. 1408.

\bibitem{Gurney2005}
Gurney P, Pauly J, Nishimura DG.
\newblock {A Simple Method for Measuring B0 Eddy Currents}.
\newblock Proc Int Soc Magn Reson Med 2005;\hspace{0pt}p. 866.

\end{thebibliography}

\end{document}